\documentclass[11pt,a4paper]{article}
\usepackage[utf8]{inputenc}
\usepackage{graphicx}
\usepackage{amsmath,amssymb,mathtools}
\usepackage{hyperref}
\hypersetup{colorlinks=true,linkcolor=blue,citecolor=blue,urlcolor=blue}
\usepackage{geometry}
\usepackage{multirow}
\date{}

\begin{document}

\title{{\bf Relativistic transformation of temperature for exotic cosmological fluids}}

\author{Soroor Pouryazdan\thanks{%
e-mail: s.pouryazdanpanahkermani@iau.ac.ir}\,\, and Babak Vakili\thanks{email:
ba.vakili@iau.ac.ir (corresponding author)}\\\\{\small {\it
Department of Physics, CT.C., Islamic Azad
University, Tehran, Iran}}} \maketitle

\begin{abstract}
We study the relativistic transformation of temperature for effective
cosmological fluids with negative pressure within a covariant thermodynamic
framework. The Lorentz transformation of temperature is derived for barotropic
fluids with $p=w\rho$ and for the (generalized) Chaplygin gas with
$p=-A/\rho^\alpha$. For barotropic fluids with constant negative
equation-of-state parameter, we show that negative pressure suppresses the
growth of temperature under relativistic boosts. In contrast, the Chaplygin gas
exhibits a qualitatively different behavior due to its density-dependent
pressure, leading to a monotonic increase of the transformed temperature with
velocity. The dependence of the relativistic temperature on the relative
velocity and background energy density is illustrated through representative
examples.
\vspace{5mm}\noindent\\
Keywords: Relativistic temperature transformation; Barotropic fluids; Chaplygin gas; Negative pressure; Cosmological fluids; Lorentz boosts

\end{abstract}

\section{Introduction}

The concept of temperature in relativistic systems has been a subject of
long-standing debate, with foundational work dating back to Planck and
Einstein~\cite{Einstein1907, Planck1908}. The question of how temperature
transforms under Lorentz boosts has important implications for both
fundamental thermodynamics and astrophysical or cosmological applications and the issue has accompanied
the development of relativistic thermodynamics since the early years of special relativity \cite{tolman1934}.
Despite more than a century of discussion, the problem of transformation laws of temperature remain mutually inconsistent,
and the correct relativistic behavior of temperature continues to be debated in both classical and quantum contexts
\cite{vanKampen1968, Israel1976, Costa1995, Cubero2007}. While the transformation of temperature for ideal relativistic gases, such as
photons and Maxwell--J\"uttner distributions~\cite{juttner1911},
is now relatively well understood, see our previous study \cite{Vakili1}, the extension to exotic or effective
cosmological fluids remains largely unexplored.

In modern cosmology, a variety of fluid models with negative pressure have
been employed to describe dark energy, accelerated expansion or unified
dark matter-dark energy scenarios~\cite{riess1998,perlmutter1999,kamenshchik2001}.
Among these, barotropic fluids with constant negative equation-of-state (EoS)
parameter $w<0$ provide a simple effective description, while Chaplygin and
generalized Chaplygin gases~\cite{bento2002,gorini2003} introduce a
density-dependent negative pressure leading to a nontrivial cosmological
dynamics. Despite their widespread use in cosmology, the relativistic
thermodynamics of such fluids, particularly under Lorentz transformations,
has not been systematically analyzed.

In this work, we address this gap by studying the relativistic transformation
of temperature for barotropic fluids with negative $w$ and for Chaplygin-type
gases. Using a covariant thermodynamic formalism, we derive explicit
expressions for the transformed temperature and identify the role of the
enthalpy density and the equation of state. For barotropic fluids, we find
that negative pressure suppresses the growth of temperature under boosts,
while for the Chaplygin gas the transformation law exhibits a qualitatively
different, monotonically increasing behavior due to the density-dependent
pressure. Representative examples are illustrated through plots of $T'/T$
as a function of the relative velocity, highlighting the interplay between
Lorentz kinematics and exotic equations of state. These results provide new
insights into the relativistic thermodynamic behavior of effective cosmological
fluids and pave the way for future studies in accelerated frames or
curved spacetime backgrounds.

\section{Relativistic temperature: operational definition and Lorentz transformation}

Before addressing cosmological fluids with exotic EoS, it is essential
to briefly recall the operational meaning of temperature in relativistic systems and
its transformation properties under Lorentz boosts. More recent studies have pointed out that no single transformation rule can be universally valid,
since the notion of temperature depends on the measurement procedure: whether it is inferred from a thermometer in thermal contact, from the shape of the radiation spectrum,
or from the energy–momentum tensor of a moving medium.
Hence, the temperature of a moving system may not be a unique scalar quantity,
but an observer-dependent concept \cite{mares2017}-\cite{hao2024}.
Our approach follows the standpoint that temperature is not a fundamental Lorentz scalar,
but a derived thermodynamic quantity whose transformation law depends on the physical
definition adopted and on the nature of the underlying system.

Following the covariant kinetic formulation of relativistic fluids, the relations
between microscopic distribution functions and macroscopic quantities
are obtained in the standard way \cite{Hakim1968, Hakim 2}. In special relativity, all macroscopic thermodynamic quantities of a continuous medium
are encoded in the energy–momentum tensor 

\begin{equation}
T^{\mu\nu} = (\rho + p) u^\mu u^\nu + p \eta^{\mu\nu},
\label{emtensor}
\end{equation}
where $\rho$ and $p$ denote the energy density and pressure in the rest frame of the fluid,
$u^{\mu}$ is the four-velocity satisfying the  normalization condition $u^\mu u_\mu = -1$, which ensures that $\rho$ is uniquely defined
as the energy density in the local rest frame, and $\eta^{\mu\nu} = \mathrm{diag}(-1,1,1,1)$
is the Minkowski metric (we use units with $c=1$). 

In cosmology and relativistic hydrodynamics, a wide class of physical systems
is effectively described in terms of perfect fluids, regardless of the existence
of a well-defined microscopic particle interpretation. Such an effective description proves particularly useful for modeling dark-energy-like
components and exotic fluids characterized by unconventional equations of state. In writing (\ref{emtensor}), no assumption is made regarding the microscopic origin of the fluid;
the formalism applies equally to conventional relativistic gases
and to effective cosmological fluids. For a perfect fluid at rest, $u^{\mu} = (1,0,0,0)$ and therefore

\begin{equation}
T^{\mu\nu}_{\text{rest}} = \mathrm{diag}(\rho,p,p,p).
\end{equation}
When the system moves with a constant velocity $v$ along the $x$-direction
relative to an inertial observer, its four-velocity becomes
$u^{\mu} = \gamma(1,v,0,0)$.
Under a Lorentz boost $x'^{\mu}=\Lambda^{\mu}_{\hspace{2mm}\nu}x^{\nu}$, with

\begin{equation}
\Lambda^{\mu}_{\hspace{2mm}\nu}=
\begin{pmatrix}
\gamma & -\gamma v & 0 & 0 \\
-\gamma v& \gamma & 0 & 0 \\
0 & 0 & 1 & 0 \\
0 & 0 & 0 & 1
\end{pmatrix},
\end{equation}
equation (\ref{emtensor}) yields in the moving frame $S'$:

\begin{eqnarray}\label{eq:rho_prime}
\left\{
\begin{array}{ll}
T'^{00}= \gamma^{2}(\rho + p v^{2}),\\\\
T'^{0x}= -\gamma^{2} v (\rho + p),\\\\
T'^{xx}= \gamma^{2}(v^{2} \rho + p).
\end{array}
\right.
\end{eqnarray}
It is important to emphasize that the Lorentz transformation acts on the components
of the energy-momentum tensor, not directly on temperature.
Any transformation law for $T$ must therefore be inferred indirectly,
through the thermodynamic relations connecting $T$ to $\rho$, $p$ and entropy density $s$.
Indeed, since temperature is defined through derivatives of thermodynamic potentials,
its transformation under Lorentz boosts is not universal.
Instead, it depends on the EoS and on the physical system under consideration.
Our previous study has shown that for specific systems such as black body radiation,
relativistic Maxwell--J\"uttner and electron gases,
the temperature transforms in a well-defined manner
once the operational definition is fixed \cite{Vakili1}. In the present work, we adopt the same operational viewpoint and extend it to
effective cosmological fluids with unconventional EoS. In particular, we shall demonstrate that while the Lorentz transformation of
energy density and pressure is purely kinematical, the resulting transformation law for temperature is sensitive to the EoS,
and may exhibit qualitatively new features for fluids with negative pressure.

In relativistic thermodynamics, temperature is most naturally defined through its
thermodynamic relation to entropy and energy. In a local equilibrium state, one may define
the inverse temperature as

\begin{equation}
\frac{1}{T} \equiv \left(\frac{\partial S}{\partial E}\right)_{V,N},
\end{equation}
where $S$ denotes the entropy and $E$ the total energy measured in a given reference frame.
This definition is purely operational and does not rely on any microscopic interpretation
in terms of particle degrees of freedom. For a relativistic fluid, it is convenient to work with local densities.
The first law of thermodynamics can be written as

\begin{equation}
d\rho =nT\,d(s/n) + (\rho + p)\frac{dn}{n},
\label{firstlaw}
\end{equation}
where $\rho$ is the energy density, $p$ the pressure, $s$ the entropy density
and $n$ the particle (or effective) number density. Equation (\ref{firstlaw}) remains valid for any effective perfect fluid,
including those without a well-defined microscopic particle content, such as cosmological fluids with negative pressure. For cosmological fluids, $n$ should be regarded as a bookkeeping variable
ensuring the validity of the thermodynamic relations, rather than a physical particle number density. Assuming local thermodynamic equilibrium and adiabatic flow,
the entropy per effective particle

\begin{equation}
\sigma \equiv \frac{s}{n},
\end{equation}
is conserved along the fluid worldlines. This condition implies

\begin{equation}
d\sigma = 0,
\end{equation}
which plays a central role in determining the temperature dependence
on the energy density.

The thermodynamic properties of the fluid are closed by specifying
an EoS relating pressure and energy density

\begin{equation}
p = p(\rho).
\end{equation}
Within the effective fluid description, temperature is not introduced
as a fundamental variable but emerges from the thermodynamic relations.
Combining the EoS with the first law and the adiabaticity condition,
one can express the temperature as a function of the energy density

\begin{equation}
T = T(\rho).
\end{equation}
This viewpoint is particularly advantageous for exotic cosmological fluids,
for which a microscopic definition of temperature may be ambiguous or unavailable.
In the following sections, we shall exploit this formalism to derive
the Lorentz transformation properties of temperature for barotropic fluids
with negative pressure and Chaplygin-type EoS, commonly employed in cosmology.

\section{Barotropic fluids with negative pressure}

We begin our analysis with barotropic relativistic fluids characterized by a linear EoS

\begin{equation}
p = w \rho,
\label{eosw}
\end{equation}
where the EoS parameter $w$ is assumed to be constant.
While this class includes standard relativistic matter and radiation,
our main interest lies in the regime $w<0$, which is commonly employed
as an effective description of dark-energy-like components in cosmology. This equation together with the first law of thermodynamics (\ref{firstlaw}),
allows one to determine the functional dependence of temperature on the energy density. Introducing the entropy per effective particle $\sigma = s/n$ and imposing
$d\sigma = 0$, the first law reduces to

\begin{equation}\label{A}
d\rho = (\rho + p)\frac{dn}{n}.
\end{equation}
For a barotropic fluid this yields

\begin{equation}
\frac{d\rho}{\rho} = (1+w)\frac{dn}{n},
\end{equation}
which can be readily integrated to give

\begin{equation}
\rho \propto n^{1+w}.
\label{rhon}
\end{equation}
On the other hand, in relativistic thermodynamics, the enthalpy density is defined as
$H \equiv \rho + p$, and the enthalpy per particle is given by
$h = (\rho+p)/n$. For an isentropic fluid with constant entropy per particle,
the temperature is proportional to the enthalpy per particle \cite{landau1980}, leading to

\begin{equation}
T \propto \frac{\rho + p}{n}
= (1+w)\frac{\rho}{n}.
\end{equation}
Combining this relation with (\ref{rhon}) leads to

\begin{equation}
T \propto \rho^{\frac{w}{1+w}},
\label{Trho}
\end{equation}
which holds for all $w \neq -1$.
Equation (\ref{Trho}) shows that temperature is a derived quantity
whose dependence on energy density becomes qualitatively different
for fluids with negative pressure.

We now consider an inertial observer moving with constant velocity $v$
relative to the comoving frame of the fluid.
As mentioned in Section~2, the Lorentz transformation acts on the components of the energy-momentum tensor.
For a boost along the $x$-direction, the energy density measured by the moving observer is given by (\ref{eq:rho_prime}) with the result 

\begin{equation}
\rho' = \gamma^2 \rho (1 + w v^2).
\label{rhoprime}
\end{equation}
This transformation is purely kinematical and remains valid independently
of the thermodynamic interpretation of the fluid. Since temperature is a function of the energy density,
its transformation law follows indirectly from Eqs.~(\ref{Trho}) and (\ref{rhoprime}). Substituting $\rho'$ into (\ref{Trho}), we obtain

\begin{equation}
T' = T
\left[\gamma^2 (1 + w v^2)\right]^{\frac{w}{1+w}}.
\label{Tprime}
\end{equation}
Equation (\ref{Tprime}) constitutes the main result of this section.
It shows explicitly that the relativistic transformation of temperature
depends on the EoS parameter $w$.
For standard fluids, known results are recovered: In the case of radiation ($w=1/3$), one finds $T' = T \gamma^{1/2}(1+v^2/3)^{1/4}$,
in agreement with the relativistic photon gas \cite{Vakili1}, and for pressureless matter ($w=0$), temperature remains invariant,
reflecting the absence of a meaningful thermodynamic temperature in this limit.

\begin{figure}[h!]
\centering
\includegraphics[width=0.5\textwidth]{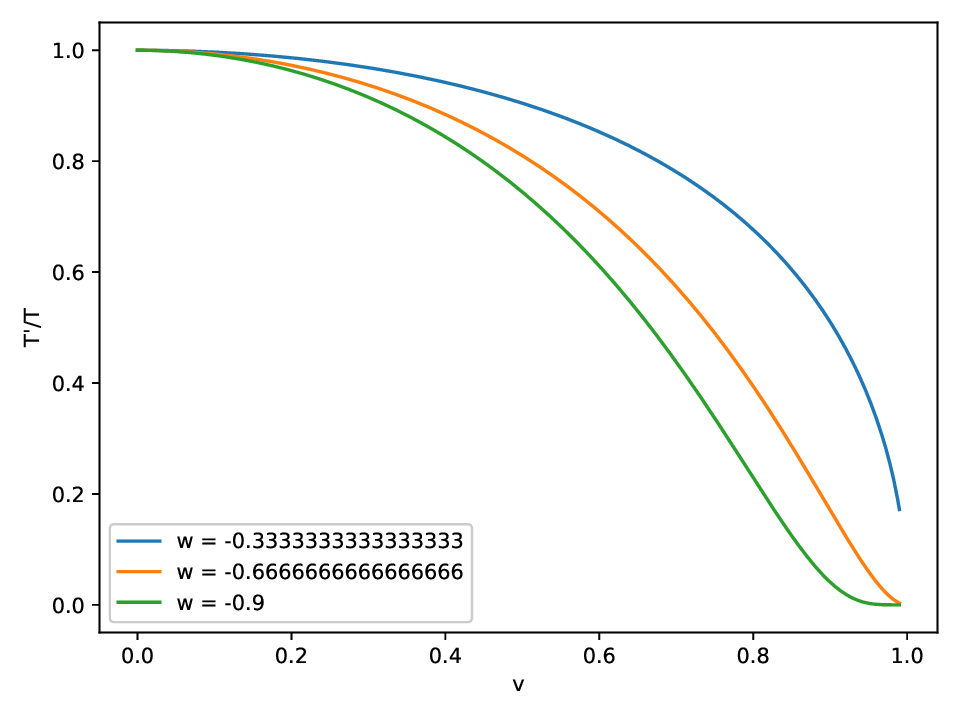}
\caption {Relativistic temperature ratio $T'/T$ versus the relative velocity $v$
for barotropic fluids with negative EoS parameter.
Curves correspond to $w=-1/3$, $w=-2/3$, and $w=-0.9$.}
\label{fig1}
\end{figure}

For fluids with negative pressure ($w<0$),
Eq.~(\ref{Tprime}) predicts a qualitatively different behavior.
In particular, the temperature may decrease under Lorentz boosts,
a feature that has no analogue in conventional relativistic gases. The case $w \to -1$ corresponds to a cosmological-constant-like fluid,
for which $\rho + p = 0$. In this limit, the notion of temperature becomes ill-defined, as already anticipated from the vanishing enthalpy density.
This behavior is fully consistent with the interpretation of a cosmological constant as a non-thermal vacuum component. For $-1 < w < 0$, the temperature remains well defined
but exhibits a nontrivial frame dependence. This result does not imply any violation of relativistic covariance,
but rather reflects the derived nature of temperature and its sensitivity to the equation of state. Figure~\ref{fig1} illustrates the relativistic transformation of
temperature for barotropic fluids with negative pressure by showing the ratio
$T'/T$ as a function of the relative velocity $v$ for several representative
values of the EoS parameter $w<0$. In contrast to ordinary
relativistic gases with positive pressure, the presence of negative pressure
introduces a nontrivial modification in the velocity dependence of the
transformed temperature. While the Lorentz factor $\gamma$ monotonically
increases with $v$, its effect is partially counterbalanced by the pressure
contribution encoded in the factor $(1+w v^2)$. As a consequence, the growth of
$T'/T$ with velocity becomes significantly milder for increasingly negative
values of $w$, and tends toward saturation as $w$ approaches $-1$. This behavior
reflects the intrinsically exotic thermodynamic response of cosmological fluids
with negative pressure under relativistic boosts.

These features highlight the fact that relativistic temperature transformation
is not universal, but encodes valuable information about the underlying
thermodynamic structure of the fluid.

\section{Chaplygin gas}

We now extend our analysis to Chaplygin-type fluids, which are characterized by
a nonlinear equation of state and are widely used as effective models interpolating between matter-dominated and dark-energy-dominated regimes
in cosmology. The (generalized) Chaplygin gas is defined by the EoS

\begin{equation}
p = -\frac{A}{\rho^\alpha},
\label{chap_eos}
\end{equation}
where $A>0$ and $\alpha \geq 0$ are constants.
The original Chaplygin gas corresponds to $\alpha=1$. As in the previous sections, we treat the Chaplygin gas as an effective
relativistic perfect fluid and assume local thermodynamic equilibrium and adiabatic flow. No microscopic particle interpretation is required for the validity
of the thermodynamic relations. Under the adiabaticity condition $d\sigma = 0$,
the first law (\ref{A}) reduces to

\begin{equation}
\frac{d\rho}{dn}
= \frac{1}{n}\left(\rho - \frac{A}{\rho^\alpha}\right).
\end{equation}
This equation can be integrated to yield

\begin{equation}
\rho^{\alpha+1} = A + B\, n^{\alpha+1},
\label{chap_rho_n}
\end{equation}
where $B$ is an integration constant. This equation explicitly shows how the Chaplygin gas interpolates
between a dust-like regime at high energy density and a cosmological-constant-like regime at low energy density.

The temperature follows from the thermodynamic identity

\begin{equation}
T \propto \frac{\rho + p}{n}
= \frac{1}{n}\left(\rho - \frac{A}{\rho^\alpha}\right).
\label{chap_Tdef}
\end{equation}
Using Eq.~(\ref{chap_rho_n}) to eliminate $n$, we find

\begin{equation}
T \propto
\left(\rho^{\alpha+1} - A\right)^{\frac{\alpha}{\alpha+1}}
\, \rho^{-\alpha}.
\label{chap_Trho}
\end{equation}
Equation (\ref{chap_Trho}) shows that the temperature is a highly nonlinear
function of the energy density. In the high-density regime $\rho^{\alpha+1} \gg A$, the pressure term becomes negligible, $p \simeq 0$, and Eq.~(\ref{chap_rho_n}) reduces to

\begin{equation}
\rho^{\alpha+1} \simeq B\, n^{\alpha+1},
\end{equation}
implying $\rho \propto n$. As a consequence, Eq.~(\ref{chap_Tdef}) yields

\begin{equation}
T \propto \frac{\rho}{n} \simeq \text{const}.
\end{equation}
This result reflects the dust-like nature of the Chaplygin gas in the
high-density regime, where a meaningful thermodynamic temperature
ceases to play a dynamical role. In contrast, in the low-density limit $\rho^{\alpha+1} \to A$, one finds $\rho + p \to 0$, and hence $T \to 0$. This behavior is consistent with the interpretation of the Chaplygin gas as approaching a cosmological-constant-like component at low energy density. However, this limit should be understood as a
breakdown of the thermodynamic description rather than the existence of a
physical zero-temperature fluid, since a cosmological constant corresponds
to vacuum and does not admit a meaningful thermodynamic temperature.

For an observer moving with velocity $v$ relative to the comoving frame, the energy density transforms according to

\begin{equation}
\rho' = \gamma^2 \left(\rho + p v^2\right)
= \gamma^2 \left(\rho - \frac{A v^2}{\rho^\alpha}\right).
\label{chap_rhoprime}
\end{equation}
Since the temperature is a function of the energy density, its transformation law follows by replacing $\rho$ with $\rho'$
in Eq.~(\ref{chap_Trho}). This yields

\begin{equation}
T' \propto
\left[(\rho')^{\alpha+1} - A\right]^{\frac{\alpha}{\alpha+1}}
(\rho')^{-\alpha},
\label{chap_Tprime}
\end{equation}
with $\rho'$ given by Eq.~(\ref{chap_rhoprime}). Unlike the barotropic case, the transformation of temperature
for the Chaplygin gas cannot be expressed as a simple power of the Lorentz factor.
Instead, it exhibits a nonlinear dependence on both the boost velocity
and the energy density, reflecting the intrinsically nonlinear equation of state. Also, it is seen that the vanishing of temperature in the limit $\rho^{\alpha+1} \to A$
is a robust feature of Chaplygin-type fluids and is independent of the observer's state of motion.
This behavior is consistent with the interpretation of the Chaplygin gas as interpolating toward a cosmological-constant-like component at late times.

\begin{figure}[h!]
\centering
\includegraphics[width=0.5\textwidth]{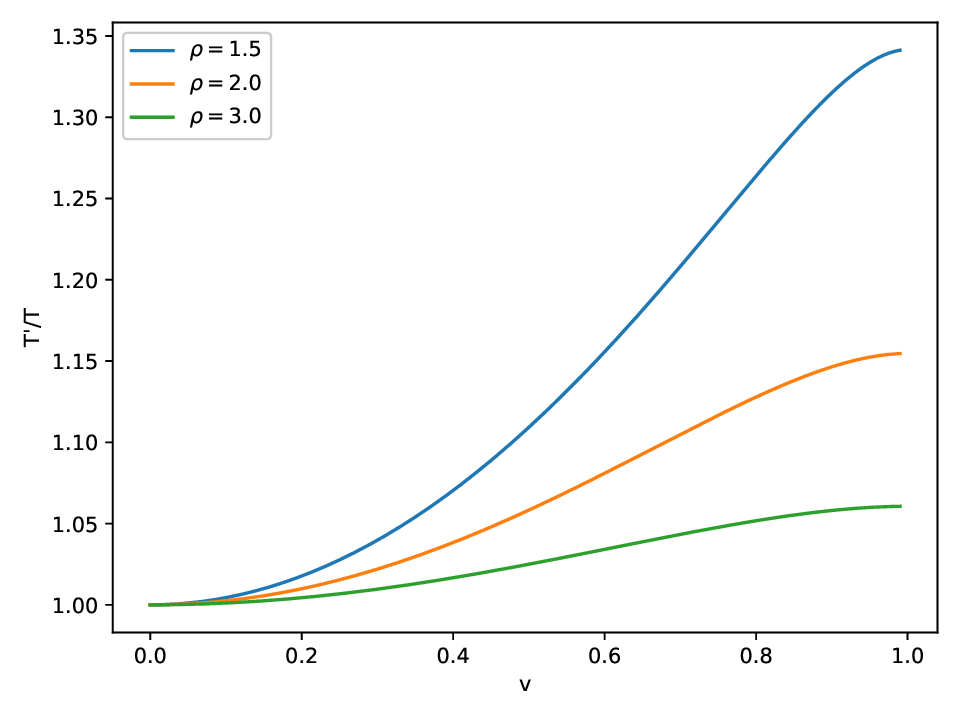}
\caption {Relativistic temperature ratio $T'/T$ as a function of the relative velocity $v$
for the Chaplygin gas with $\alpha=1$.
Different curves correspond to different background energy densities $\rho$,
illustrating the density-dependent nature of the temperature transformation.}
\label{fig2}
\end{figure}

Figure~\ref{fig2} displays the relativistic transformation of temperature
for the Chaplygin gas by plotting the ratio $T'/T$ as a function of the relative
velocity $v$ for several representative background energy densities $\rho$.
In contrast to barotropic fluids with a fixed negative EoS
parameter, the transformed temperature in the Chaplygin model exhibits a
monotonically increasing behavior with velocity. This feature originates from
the density-dependent nature of the Chaplygin pressure,
$p=-A/\rho^\alpha$, which leads to an effective EoS parameter
$w_{\rm eff}(\rho)=-A/\rho^{\alpha+1}$ that becomes progressively less negative
as the boosted energy density $\rho'$ increases. Consequently, the Lorentz
enhancement of the energy density dominates over the negative-pressure
contribution, resulting in an overall growth of the transformed temperature.
The dependence of the curves on the background density further highlights the
non-barotropic character of the Chaplygin gas and distinguishes its
thermodynamic response from that of ordinary and exotic barotropic fluids.

Again, the nontrivial frame dependence of temperature for intermediate regimes
does not signal any violation of relativistic covariance,
but rather highlights the derived nature of temperature
and its sensitivity to the thermodynamic structure of the fluid. These results further support the conclusion that relativistic
temperature transformation laws are not universal, but encode detailed information about the underlying equation of state.

\section{Conclusion}

In this work, we have investigated the relativistic transformation of temperature
for effective cosmological fluids with negative pressure, focusing on two
representative classes: barotropic fluids with constant negative EoS
parameter $w$ and the (generalized) Chaplygin gas with $p=-A/\rho^\alpha$. Using
a covariant thermodynamic formalism\footnote{Although our approach does not rely on a fully covariant entropy current
formalism, it consistently treats thermodynamic quantities as local scalar
fields and ensures Lorentz covariance of the resulting temperature
transformation laws.}, we derived explicit expressions for the
transformed temperature under Lorentz boosts and examined their dependence on
the relative velocity and the background energy density.

A key quantity underlying all cases studied in this work is the enthalpy density,
$\rho + p$. For relativistic fluids in local equilibrium, temperature naturally emerges through
the combination $(\rho + p)/n$, independently of any microscopic interpretation.
As a consequence, the behavior of temperature under Lorentz boosts is ultimately
controlled by how $\rho + p$ transforms between inertial frames. In the case of a barotropic fluids with $p = w\rho$, the enthalpy density scales linearly with
the energy density, leading to a simple power-law relation between temperature
and $\rho$, and consequently to an explicit transformation law for $T$ under
Lorentz boosts. On the other hand, for Chaplygin-type fluids, the nonlinear structure of the EoS gives rise to a much richer behavior, where no simple scaling with the Lorentz factor exists.

Although both barotropic fluids with $w<0$ and Chaplygin-type fluids are commonly
used as effective models of dark energy, their thermodynamic behavior differs
in an essential way. For barotropic fluids, we find that the negative pressure acts to suppress the
growth of temperature with increasing velocity, in agreement with the
expected effect of a fixed negative EoS parameter. In contrast,
the Chaplygin gas exhibits a qualitatively different behavior: the density-dependent
nature of its pressure leads to a monotonic increase of the transformed temperature
with velocity. This growth is further modulated by the background energy density,
highlighting the non-barotropic character of the Chaplygin gas and the interplay
between Lorentz kinematics and exotic equations of state. At high energy density, the fluid behaves as pressureless matter and the temperature
approaches a constant value, reflecting the dust-like nature of the system.
At low energy density, the fluid asymptotically approaches a
cosmological-constant-like state, for which the temperature vanishes.
In intermediate regimes, the transformation of temperature under Lorentz boosts
exhibits a genuinely nonlinear dependence on both the velocity and the energy density.

Representative examples of $T'/T$ as functions of the relative velocity
illustrate these behaviors and provide a clear comparison between the two classes
of fluids. The results emphasize that the thermodynamic response of cosmological
fluids under relativistic boosts is not solely determined by the sign of the
pressure, but also by its density dependence. In summary, the obtained results provide a unified picture of
relativistic temperature transformations for a broad class of effective
cosmological fluids. A central outcome of our analysis is that the transformation law of temperature
is not universal, but depends sensitively on the thermodynamic structure of the fluid, in particular on its EoS. However, the frame dependence of temperature found in this work should be interpreted with care. Temperature is not a fundamental Lorentz scalar but a macroscopic quantity
defined through thermodynamic relations. Its transformation properties therefore depend on the physical assumptions
underlying the effective fluid description, such as local equilibrium and adiabaticity.

Overall, our findings extend the understanding of relativistic thermodynamics
to fluids commonly used in cosmology and underscore the importance of considering
the equation of state in analyses of temperature transformations. These insights
pave the way for future investigations of more complex scenarios, such as
relativistic fluids in accelerated frames or in curved spacetime backgrounds,
and may have implications for the thermodynamic description of dark energy
and unified dark matter-dark energy models.

\end{document}